\documentclass[floatfix,twocolumn,prb,aps,epsf,amsmath,amssymb,showpacs]{revtex4-1}
\usepackage{graphicx}
\usepackage{comment}

\begin{document}

\preprint{in progress}

\title
{Clausius Implies That Nearly Anything Can Be A Thermometer}
\date{\today}
\author{Wayne M. Saslow,$^{1}$ }\email{wsaslow@tamu.edu}
 \affiliation{
$^{1}$ Department of Physics, Texas A\&M University, College Station, Texas, 77843-4242, U.S.A.,  \\
 }

\begin{abstract}

There are three types of thermometries.  One is a proxy, such as the purely phenomenological resistivity.  More fundamental are those based on statistical mechanics, as with the ideal gas law, and those based on thermodynamics, as with the basic Carnot cycle.  With heat input $Q$ and temperature $T$, in principle (but not in practice) a temperature scale can be based on the basic Carnot cycle relation $Q/T+Q'/T'=0$, with a temperature $T_{0}(p_{0},V_{0})$ specified.   More generally, a thermodynamics based temperature scale can be determined by 
the Clausius condition $\oint dQ/T=0$ for every closed path in a given region $\Omega$ of $p$-$V$ space.  Taking a discretized grid $i$ (from which such closed paths can be composed), for some parametrized model temperature function $T_{n}$ 
a root-mean-square minimization of $\sum_{i}(\oint_{i}dQ/T_{n})^{2}$ yields the best set of model $T_{n}$'s parameters.   Thus any stable material -- even one not described by a statistical mechanical model -- can be used as a thermometer.  
If, because of inaccuracy of $dQ$ measurement, the rms Clausius condition temperature scale gives lower accuracy than the best statistical mechanics temperature scale, then that statistical mechanics temperature scale can be employed with the rms Clausius condition approach to improve the accuracy of (i.e., raise the standards for) the $dQ$ measurements to the accuracy of the statistical mechanics based temperature scale.  

\end{abstract}

\maketitle

\section{Introduction}

It is well-known that the basic four-leg Carnot cycle condition\cite{Sears,Callen1,Fermi,Reif}  
for two thermal reservoirs and reversible processes relates the heat inputs $Q$ into each reservoir and their temperatures $T$ by 
\begin{equation}
\frac{Q}{T}+\frac{Q'}{T'}=0.  \qquad \hbox{(basic Carnot)}
\label{Carnotsimple}
\end{equation}
In principle, if some $T$ is fixed -- such as at $T_{0}(p_{0},V_{0})$ -- and $Q$ and $Q'$ are measured (entering is taken to be positive), then $T'$ can be determined.  This gives a fundamental thermodynamic definition of temperature.  However, if $T$ and $Q$ are known, then $Q'$ must be adjusted in order to find the appropriate Carnot cycle.  Moreover, enforcing zero heat input along two legs and constant temperature along the other two legs may impose difficult practical constraints.  We know of no implementation of the Carnot cycle procedure in practical thermometry.  %so this likely is an impractical approach.  

In practice, thermometry is performed either with thermometers based on statistical mechanics or on proxies (such as electrical resistivity) that are then calibrated against statistical mechanics based thermometers.  Such statistical mechanics based thermometers are fundamental, and therefore are called thermodynamic temperature scales, but they require materials that can be accurately modeled in the region of interest.  This can limit their applicability. 

The present work notes that a thermodynamics based thermometry %alternate to simple Carnot cycles, 
that uses a root-mean-square fit to a general, non-Carnot, thermodynamic cycle, is not limited to materials that can be accurately modeled in the region of interest.  It therefore may provide a means to perform thermometry in previously inaccessible regions.  

Moreover, if for a given material this approach gives a temperature scale of lower accuracy than statistical mechanics based thermometry (e.g., because of low accuracy in $dQ$ measurements), then the statistical mechanics based thermometry can be used to raise the accuracy of heat measurements to the accuracy of the statistical mechanics based temperature scale.  

\begin{comment}
\section{Temporary}

What follows clarifies the basis for a previous work on thermometry, based on measurements of heat input $dQ$  for some region $\Omega$ of $p$-$V$ space, and the Clausius condition that $\oint dQ/T=0$ for all paths in $\Omega$.\cite{SaslowEJP20}  The present work extends the previous work by: recognizing that there are three categories of thermometries, only one of which is truly thermodynamics-based, and by definition not a proxy; considering the implications for thermometry of a discretization of $p$-$V$ space; and emphasizing that all thermometries have their scale determined by some known temperature $T_{0}(p_{0},V_{0})$ (such as at a triple point) within $\Omega$.  

The three types of thermometries are, in addition to the non-proxy thermodynamics-based temperature scale considered here, which apply to any material, temperature proxies that either are purely phenomenological (such as the electrical resistivity), or are based on statistical mechanics (such as the ideal gas law).  This will be expanded on below. 
If the Clausius-based approach for a given material gives a temperature scale of lower accuracy than a proxy, then in the traditional spirit of standards laboratories, that proxy can be employed to raise the accuracy of the $dQ$ measurements to the accuracy of the proxy temperature scale.  

If a laboratory is without any thermometer at all, then any available material (Fe, Cu, NaCl, etc.) can be turned into a thermometer using the Clausius-based approach.  
\end{comment}

\section{Thermometry using the Clausius condition}

\begin{comment}
The basic Carnot cycle for two reservoirs and reversible processes relates the heat input into each reservoir and their temperatures by 
\begin{equation}
\frac{Q_{1}}{T_{1}}+\frac{Q_{2}}{T_{2}}=0.  \qquad \hbox{(Carnot cycle)}
\label{Carnot1}
\end{equation}
In principle, if $T_{1}$ is fixed -- such as at $T_{0}(p_{0},V_{0})$ -- and $Q_{1}$ and $Q_{2}$ are measured (entering is taken to be positive), then $T_{2}$ can be determined.  This gives a thermodynamic definition of temperature.  However, if $T_{1}$ and $Q_{1}$ are fixed, $Q_{2}$ must be adjusted in order to find the appropriate Carnot cycle, so this likely is an impractical approach.  We propose using a different version of this result. 
\end{comment}

With heat input $dQ$ (taken as positive if it enters the system), Clausius showed\cite{Clausius65,Clausius54} what we will call the Clausius condition --- that for an unspecified integration factor $f\equiv1/T$ (so is $T$ unspecified), the closed-path line integral for the change in what he later called entropy $S$ is zero:  
\begin{equation}
\Delta S=\oint fdQ=\oint\frac{dQ}{T}=0. \qquad \hbox{(Clausius condition)}
\label{ClausiusDelS}
\end{equation}
With the temperature specified at some point in $p$-$V$ space, this can be used to define $T(p,V)$ --- or to constrain the $dQ$'s.  Clausius analyzed a general Carnot cycle for an ideal gas (we will take temperature symbol $T_{g}$, so $pV=nRT_{g}$).  With an implicitly fixed temperature specified to determine the ideal gas scale, he found $T=T_{g}$.  Thus this initial application found that the thermodynamic temperature corresponds to the ideal gas temperature, which is an example of a statistical mechanical temperature (for a non-interacting gas).  However, eq.~\eqref{ClausiusDelS} serves to {\it define} temperature more generally, in the sense of thermodynamics applied to any system.  All systems must give the same thermometry, although their equation of state --- and therefore their entropy $S$ --- can be expected to differ. 

\begin{figure}[!htb]
\hspace{-1.59cm}
%\begin{center}
\vspace{-0.8cm}
\hspace{0.6cm}
\includegraphics[width=2.8in]{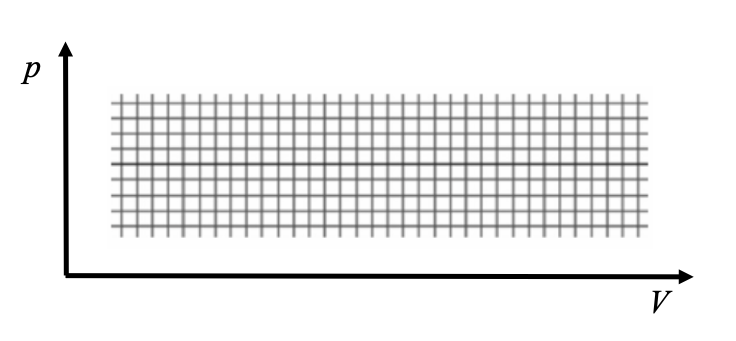}
\vspace{0.2cm}
\footnotesize {\caption{\label{fig:pvgrid} Possible discretization in $p$-$V$ space for use in testing different proposed temperature scales.}  }
\end{figure}

For systems more complex than an ideal gas, the general Carnot cycle can only be evaluated numerically.\cite{Carnotsimple}  To try to implement \eqref{ClausiusDelS}, imagine that the full region $\Omega$ is discretized into $i$ polygons, as in Fig.~\ref{fig:pvgrid}.  For a correctly chosen temperature, \eqref{ClausiusDelS} must be satisfied for each polygon.  Written for each $i$, and with all relevant $dQ$ known, the terms in $1/T_{0}$, with $T_{0}$ known, become source terms in linear equations for the remaining unknown $1/T$.  In an $m$-by-$m$ discretization, there are $2m^{2}+2m$ measured $dQ$'s, $2m^{2}+2m$ unknown $dT$'s, and $m^{2}$ constraints.  Thus a discretization gives insufficient conditions to determine the thermodynamics-required solution for $T$.  Nevertheless, from \eqref{Carnotsimple} we know that such a solution must exist. 

\vspace{-0.01cm}

With the knowledge that fixing a single $T_{0}$ enables a meaningful temperature, the next section considers the more robust approach of satisfying \eqref{ClausiusDelS} in a root-mean-square sense, as is almost universally done in matching theory and data.  

From \eqref{ClausiusDelS}, with $T$ determined, it follows that, $dS=fdQ=dQ/T$ is a true differential, so that the entropy can be defined as\cite{Clausius65}  
\begin{equation}
S=\int\frac{dQ}{T}+\hbox{const}.
\label{entropy}
\end{equation}
The constant, of course, comes from the Third Law of Thermodynamics.  Because $S$ depends only on other thermodynamic parameters, not on the history of the thermometer, $S$ is what is known as a thermodynamic state function.\cite{Sears,Callen1,Fermi,Reif} 

\begin{comment}
If we write the energy differential $dU=dQ-dW$ as $dU=TdS-pdV$,\cite{Callen1} then
\begin{equation}
T\equiv \frac{\partial U}{\partial S}\big|_{V}.
\label{T}
\end{equation}
Thus, (\ref{ClausiusDelS}), with some $T_{0}$ specified, serves both to ensure that (a) we may define $S$ and (b) its initially unknown integration factor $f=1/T$ gives the already-known (if measurable) ideal gas temperature $T=T_{g}$.  It also means that entropy uniqueness goes hand in hand with obtaining a valid  thermodynamic temperature $T$.  
\end{comment}

\section{Root-mean-square Deviation of Entropy}

\begin{comment}
From (\ref{Clausius}) and $f\equiv 1/T$, for any reversible closed path in $p$-$V$ space one has

\begin{equation}
\Delta S=\oint\frac{dQ}{T}=0. 
\label{Clausius}
\end{equation} 
\end{comment}

We now employ eq.~\eqref{ClausiusDelS} to develop a procedure to obtain $T(p,V)$ using the root-mean-square deviation from eq.~\eqref{ClausiusDelS} for each cell $i$.\cite{SaslowEJP20}    
In Fig.~\ref{fig:pvgrid} the cell areas could be of any shape and area, but for simplicity we have taken them all to be rectangles with the same area. 

Consider a given model temperature scale $T_{n}(p,V)$, where $n$ identifies the model (including functional form and any unknown parameters); for simplicity we hereafter suppress the dependence on $p$ and $V$.  If there are $M$ such areas in the regions of interest, then the $i$-th area has a deviation 
\begin{equation}
\Delta S_{i,n}=\oint_{i} \frac{dQ}{T_{n}}. 
\label{DeltaSne0}
\end{equation}
The sign of $\Delta S_{i,n}$ has no fundamental significance, and typically will be non-zero, either because the proposed $T_{n}$ is not the true thermodynamic temperature, or because of errors associated with the discretization, or errors in the measurement of the $dQ$'s.  For each $i$, and a rectangular grid, four measurements of $dQ$ are required; a numerical approximation would take $T_{n}$ at the midpoint of each side.  

The total rms (root-mean-square) deviation, summed over the grid of $M$ areas, is given by what we will call the net rms entropic ``mismatch'' 
\begin{equation}
F_{n}=(\Delta S)^{2}_{{\rm tot},n}=\sum_{i=1}^{M}(\Delta S_{i,n})^{2}.
\label{DeltaS-rms}
\end{equation}
Of two temperature models $T_{n}$ and $T_{n'}$, the preferred model is that with the smaller $(\Delta S)^{2}_{\rm tot}$; that is, the smaller net rms entropic ``mismatch''.  

As an example, let a temperature scale have a given functional form with three parameters, one of which can be thought of as determining $T_{0}(p_{0},V_{0})$.  The best such temperature scale for this functional form is found by minimizing the rms value of (\ref{DeltaS-rms}) as a function of these three parameters for the region $\Omega$ of $(p,V)$ space.  The approach may be generalized to different functional forms and numbers of parameters.  Different regions can be stitched together to cover all of $p$-$V$ space. 

Analogously, for a $T$-$S$ diagram -- rather than a $p$-$V$ diagram -- one could employ a net rms volumetric ``mismatch'' criterion.  This would be based on the expectation that, with $dW$ the work done by the system, $\Delta V=\oint (dW/p)=0$ for each element of a thermodynamic $T$-$S$ grid.  This can provide a check on values of thermodynamic quantities. 

\section{Statistical Mechanics Based Thermometries}

Examples of thermometries based on statistical mechanics include:

(a) For a region that is accessible to an ideal, or nearly ideal, gas (with $n$ the number of moles and $R$ the gas constant), a proxy is 
\begin{equation}
pV=nRT.	
\label{idealgas}
\end{equation}
This equation works particularly well for dilute gases at somewhat elevated temperatures.  Virial expansions give corrections to eq.~\eqref{idealgas}. 

(b) A thermometry for the cosmos uses the cosmic microwave background (CMB) radiation distribution.  For frequency $\omega$ it satisfies the Bose-Einstein distribution 
\begin{equation}
n(\omega)=[e^{(\hbar\omega/k_{B}T)}-1]^{-1}.
\label{CMB}
\end{equation}  
This thermometry also is employed in modern temperature ``guns''. %also used to measure the temperature of distant objects that are significantly hotter than the earth or an extra-terrestrial observatory. 

(c) The magnetic susceptibility of a collection of nearly non-interacting nuclear spins at low temperatures (but not so low that interactions cause the nuclear spins to order) satisfies
\begin{equation}
\chi=\frac{C}{T}, 	
\label{idealmagnet}
\end{equation}
where $C$ is the Curie constant.  This is the ideal gas of magnetism. 

(d) Noise thermometry, where the noise power $P_{W}$ (watts) over a bandwidth $B$ (hertz) satisfies
\begin{equation}
\frac{P_{W}}{B}=k_{B}T,
\label{noise}
\end{equation}
where $k_{B}$ is the Boltzmann constant.  
Noise thermometry is often employed at low temperatures. 

The author was introduced to the issue of thermometry at an International Conference on Low Temperature Physics.  There two distinguished, prize-winning, low temperature physicists, who each used both susceptibility and noise thermometries, disputed which was better.\cite{SaslowEJP20}  For specific examples of practical thermometry, see Refs.~\onlinecite{Moldover,Tuoriniemi,Mazon,Casey,Saunders}.   Also note Ref.~\onlinecite{TempReview19}.

\section{Beyond Statistical Mechanics: New Types of Thermometers} 
In the absence of a thermometer or a material with properties simple enough to be described by a statistical mechanical theory, a lab must develop 
its own thermometer.  Although we earlier indicated that almost any material would serve, in practice some materials could give problems.  For example, if the material is a combination of two substances, such as air and water, then condensation or evaporation of the water might complicate the analysis.  %For thermal expansion of a metal bar, one must have an accurate way to measure its length (perhaps by interferometry).  

A possible thermometer material is an interacting gas that is too dense to be described by a statistical mechanical virial expansion.  
%Because it is non-ideal, its thermometry is not trivial, but 
However, it has the advantage that accurate pressure and volume measurements should be relatively simple to perform.  In the same spirit is an interacting set of spins,  %Because it is non-ideal, its thermometry is not trivial, but 
where accurate magnetic field, volume, and pressure measurements should be relatively simple to perform.  

In the Clausius based approach a lab can calibrate two nominally identical materials without concern that they are not identical.  In principle they would both yield the same -- or nearly the same -- temperature scale, although the $dQ$'s would differ, as would the derived entropies $S$.  

\section{Refining Measurements of Heat}
In practice, fundamental thermometry currently is performed by measuring an appropriate system, and analyzing it using a statistical mechanical theory.  %as discussed above.  
If the resulting temperature $T$ is more accurate than a corresponding Clausius based based temperature scale from an rms minimization of \eqref{DeltaS-rms} then, as is often done with standards, one can turn the problem around, and correct the heat input measurements.  

For measured heat inputs $dQ_{m}$, one uses the proxy temperature and assumes a dimensionless correction factor $F(T)$ with a few unknown parameters to obtain 
\begin{equation}
G=\sum_{i}\int_{i}\Big(\frac{dQ_{m}F(T)}{T}\Big)^{2}.
\label{rmscorrectdQ}
\end{equation}
Once $F(T)$ is determined by the minimization process, the corrected heat input $dQ_{c}$ is given by 
\begin{equation}
dQ_{c}= dQ_{m}F(T).  	
\label{correctdQ}
\end{equation}

In this way, the standards for measurement of heat input $dQ$ can be raised to the level of accuracy of the proxy-based temperature scale.  Ref.~\onlinecite{SaslowF=ma} indicates how a standard can be improved by the use of a physical law.   

It is unfortunate that there is no thermodynamic unit named after Clausius, who gave us the first correct thermodynamics,\cite{ClausiusThermo,GibbsHistoryThermo} and both introduced\cite{Clausius54} and first understood entropy.\cite{Clausius65,GibbsHistoryThermo}  A unit for entropy would surely be the clausius (Cl), rather than the (derived) joule/kelvin (J/K).  It is perhaps not coincidental that there was a priority dispute between Clausius and Kelvin.\cite{SaslowBriefHistory}\\

\section{Summary}
We have, more explicitly than previously, indicated how thermometry can be performed using a Clausius-based approach.  We observed that there are three types of thermometry, and that only the Clausius-based version is based solely on thermodynamics.  Thus it is suitable to employ nearly any material as a thermometer, not merely those for which there are statistical mechanical theories.

\section{Acknowledgements}
I would like to acknowledge valuable interactions with C. Meyer and N. Mirabolfathi.


\begin{thebibliography}{}
\bibitem{Carnotsimple} For a basic Carnot cycle involving two temperatures one has $Q/T+Q'/T'=0$.  Even if one does not know the values of $T$ and $T'$, by use of a proxy thermometer it may be possible to ensure that $T$ and $T'$ are fixed.  If one then fixes, say, $T$ at a triple point value, then measuring $Q$ and $Q'$ will then give $T'$.  This approach can be performed, but having to always ensure that the proxy thermometer is fixed may be a complicating issue in a practical sense.  On the other hand it shows that once the temperature is fixed at one point in $(p,V)$ space, it is then determined at all other points in ($p,V)$ space.  For small changes in $p$ or $V$ one must verify that $T$ and $T'$ are constant to second order in these small changes.  The integral approach we describe does not have that issue.
\bibitem{entropyname} This was established, with the symbol $N$ and without the name entropy, by Clausius in 1854; in 1865 he appreciated its significance enough to name the indefinite integral the {\it entropy}, for which the symbol $S$ was an approximative letter to its Greek source, given that $E$ was in use for {\it energy.}
\bibitem{NcalN} Clausius used the symbol $N$, but we use ${\cal N}$ to avoid confusion with particle number,
\end{thebibliography}
\end{document}